\documentclass[a4paper, 10pt, twocolumn]{article}
\pdfoutput=1
\usepackage{graphicx}
\usepackage[left=2cm,top=2cm,right=2cm,bottom=2cm,nohead]{geometry}
\usepackage[pdftex]{hyperref}
\begin{document}
\title{\large \bf Space-like Separation in a Bell Test assuming Gravitationally Induced Collapses}
\author{\normalsize D. Salart, A. Baas, J.A.W. van Houwelingen, N. Gisin, and H. Zbinden\\
\it \small Group of Applied Physics, University of Geneva, 20, Rue de l'Ecole de M\'edecine, CH-1211 Geneva 4, Switzerland}
\date{\small \today}
\twocolumn[\begin{@twocolumnfalse}
\maketitle
\begin{center}
{\small \parbox[t]{14cm}{\hspace{2mm} We report on a Bell experiment with space-like separation assuming that the measurement time is related to
gravity-induced state reduction. Two energy-time entangled photons are sent through optical fibers and
directed into unbalanced interferometers at two receiving stations separated by 18 km. At each station,
the detection of a photon triggers the displacement of a macroscopic mass. The timing ensures space-like separation from the moment a photon
enters its interferometer until the mass has moved. 2-photon interference
fringes with a visibility of up to $90.5\%$ are obtained, leading to a violation of Bell inequality.\\}}
\end{center}
\end{@twocolumnfalse}
]
When is a quantum measurement finished? Quantum theory has no definite answer to this seemingly innocent question and this leads to the
quantum measurement problem. Various interpretations of quantum physics suggest opposite views. Some state that a quantum measurement is
over as soon as the result is secured in a classical system, though without a precise characterization of classical systems. Decoherence
claims that the measurement is finished once the information is in the environment, requiring a clear cut between system and environment
and arguments assuring that the system and environment will never re-cohere. Others claim that it is never over, leading to the many worlds
interpretation \cite{manyworlds}. Note that none describes how a single event eventually happens. And there are more interpretations and
many variations on each theme. In practice this measurement problem has not yet led to experimental tests, though progress in quantum
technologies bring us steadily closer to such highly desirable tests \cite{Perci}.

Another possibility, supported among others by Penrose and Di\'osi \cite{Dio}\cite{Adl}, assumes a connection between quantum measurements and
gravity. Intuitively the idea is that the measurement process is finished as soon as space-time gets into a superposition state of
significantly different geometries. The latter would be due to superpositions of different configurations of massive objects. Penrose and
Di\'osi independently proposed the same criterion (up to a factor of 2) that relates the time of the collapse (that
terminates the measurement) to the gravitational energy of the mass distribution appearing in the superposition. Following Di\'osi's equation
\cite{Adl}, the time of the collapse is given by:

\begin{equation}
t_D=\frac{3\hbar V}{2\pi G m^2 d^2}
\end{equation}\\
\noindent
where {\it V} is the volume of the moving object, {\it m} is its mass and {\it d} is the distance it has moved.

Hence, according to Eq. 1, a typical measurement in quantum optics is finished once the alternative results would have led to
displacements of a sufficiently massive object. This view differs stridently from the one adopted in practice by most quantum opticians.
Indeed, the common view in this community is that a quantum measurement is finished as soon as the photons are absorbed by detectors.
But such an absorption, even when it triggers an avalanche photodiode and gets registered by a computer, only involves the motion of
electrons which are of insufficient mass to satisfy the Penrose-Di\'osi criterion.
\begin{center}
\includegraphics[width=1.0\linewidth]{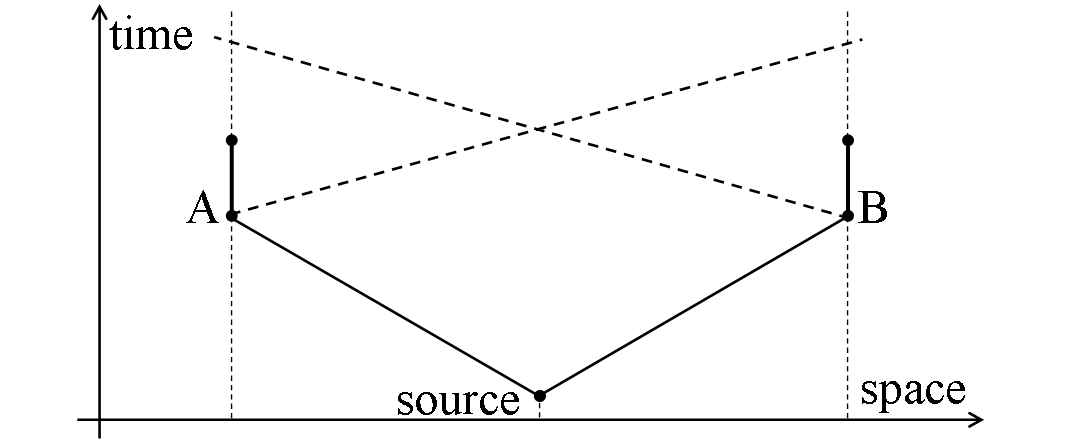}
\end{center}
{\small Fig. 1. Space-time diagram of the experiment. When a photon from the source reaches an analyzer (A or B) the measurement process starts.
Space-like separation is achieved when the measurement process at detector A (B) is finished before a light signal traveling
from A to B (B to A) arrives at the other detector.}\\

This situation led Kent to observe that actually, according to the Penrose-Di\'osi criterion, none of the many tests of Bell inequalities
that have been performed so far involve space-like separated events \cite{Kent}. Indeed, in all these tests, no massive object moves,
at least not in the microseconds following the photon absorptions by the detectors. But then, none of these Bell tests strictly excludes
the possibility that the observed violation of Bell inequalities is due to some hypothetical communication (of a type unknown to today's
physics). Given the importance of quantum nonlocality (i.e. violation of Bell inequalities), both for fundamental physics and for quantum
information science, we present in this letter an experiment that closes this loophole.

In our Franson-type test of the Bell inequalities \cite{Fran}, pairs of entangled photons traveling through optical fibers are sent to two
receiving stations physically separated by 18 km with the source at the center. This distance breaks the record for
this kind of experiment \cite{AspectNature,TitPRL}. At each receiving station, the detected photons trigger the application of a step voltage to a
piezoelectric actuator. The actuator is a ceramic-encapsulated PZT (lead zirconate titanate) block of 3x3x2 mm and weighting 140 mg
(PI, PL033). We chose this actuator because it fulfills all the following criteria: it can move a measurable distance in a time of
the order of microseconds and it can be triggered to repeat this movement several thousand times per second. Due to the converse piezoelectric
effect, the applied voltage expands the actuator and, at the same time displaces a gold-surfaced mirror measuring 3x2x0.15 mm and weighing 2 mg
that is attached to one of the piezo faces. We used this mirror as the movable mirror of a bulk optical interferometer (see Fig. 2) to confirm
the expansion of the piezo (see Fig. 3).
\begin{center}
\includegraphics[width=0.9\linewidth]{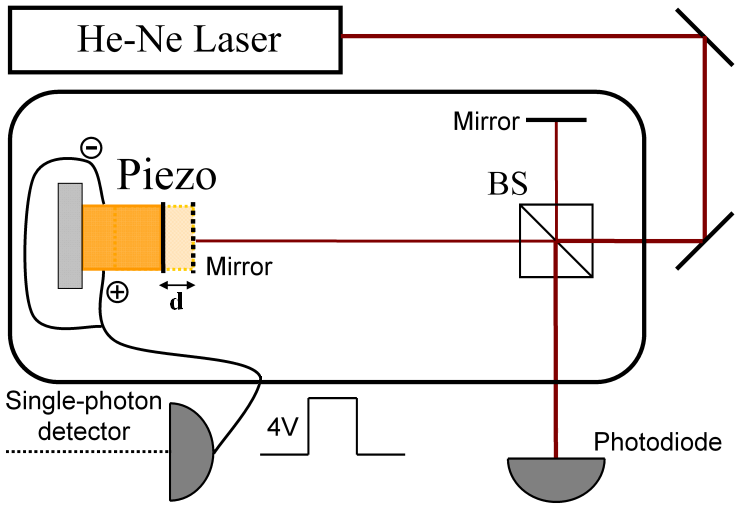}
\end{center}
{\small Fig. 2. Experimental set-up of the bulk intereferometer used in each receiving station (Satigny and Jussy) to confirm the piezo
expansion. Each interferometer is mounted inside a box that isolates it from atmospheric disturbances. The piezo actuator is glued to a
fixed support with one mirror attached to its side. Each time a photon is detected by the single-photon detector, a step voltage of 4V
is applied to the piezo, expanding it. When the piezo expands, the laser beam-path through the arm with the piezo shortens and the
interference produces a variation in the intensity observed by the photodiode.}\\

To guarantee the space-like separation between the detection events, the time that the light needs to travel from one receiving station
to the other must be significantly shorter than the time needed to perform all the measurement process ($t_M$). This time includes, not
only the time of the collapse, but also the time between the moment the photons enter their respective analyzer (a 50/50 fiber coupler
inside an interferometer), until the moment the mirrors move sufficiently to be certain that the measurement process is finished, according
to the Penrose-Di\'osi hypothesis. The time between the moment the photons enter the analyzer to the moment the step voltage is applied
is just $0.1\ \mu s$. After the application of the voltage, the piezo starts to expand and displaces the mirror. This produces a change in
the phase of the interferometer that is detected by the photodiode. The equivalence between the voltage variation detected by the
photodiode and the mirror displacement has been calculated from the wavelength of the laser ($\lambda = 633$ nm) and the phase change
produced by the displacement. If we conservatively assume that the phase change takes place in a node of an interference fringe, where the slope between
the phase and the intensity is maximum, we will set a lower bound for the displacement distance. Hence, $6\ \mu s$ after the step voltage is
applied to the piezo, the voltage has already changed by 0.3 V, for a pic-to-pic max of 2.4 V, meaning that the mirror has displaced a distance of at least 12.6 nm.  Finally,
we find that the time of collapse is $t_D= 1 \mu s$, using Eq. 1 \cite{factor2} with d=12.6 nm and taking just the mass and volume of the
mirror. The total time is then $t_M= 7.1\ \mu s$, almost one order of magnitude shorter than the $60\ \mu s$ the light needs to cover the 18 km between
the receiving stations. Note that taking into account the motion of the piezo itself would even shorten this conservative estimation of $t_M$.
\begin{center}
\includegraphics[width=0.9\linewidth]{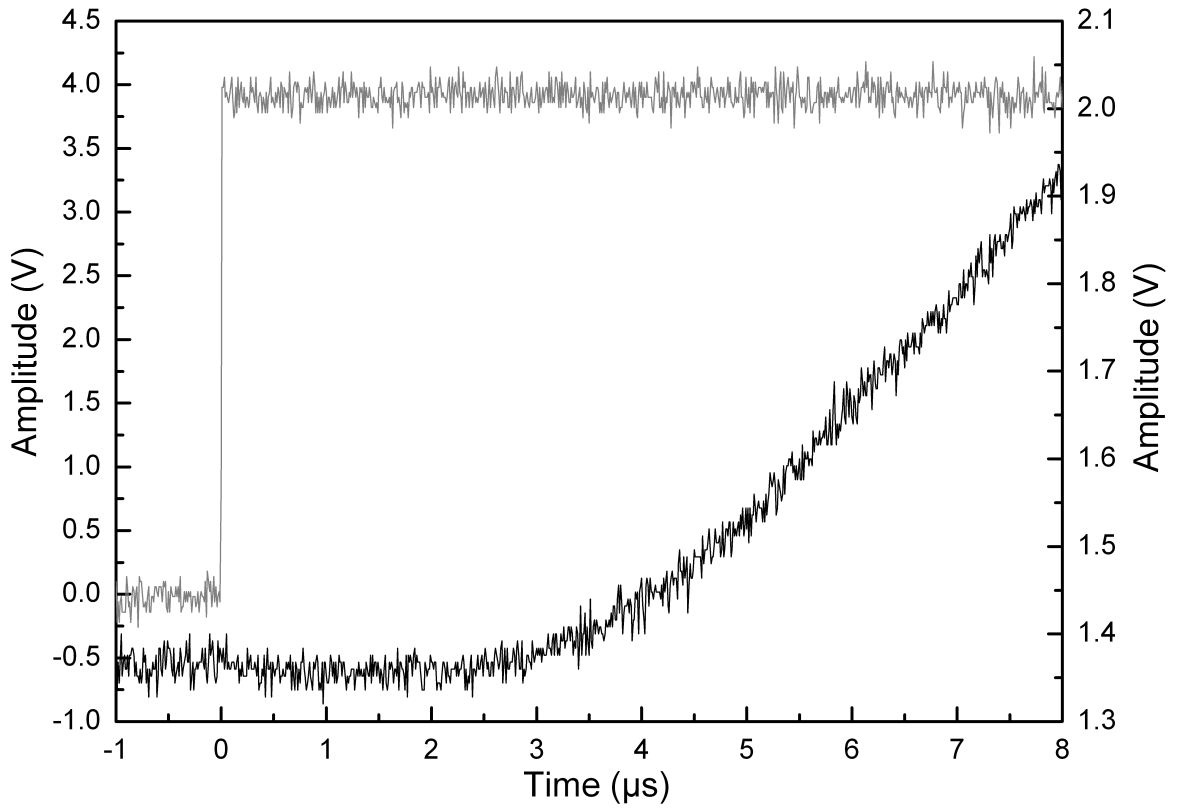}
\end{center}
{\small Fig. 3. Step voltage applied on the actuator and the mirror displacement. This measurement confirms the piezo expansion.
It was repeated in each receiving station before and after each run of the experiment. Grey line is the step voltage of 4 V
(left scale) applied to the piezo actuator. Black line is the distance the mirror has moved represented as the voltage variation (right scale)
detected by the photodiode. $6\ \mu s$ after the voltage is applied, the mirror has already moved by 12.6 nm.}\\

The scheme of the experimental setup is given in Fig. 4. A cw single mode external cavity diode laser (2.7 mW at 785.2 nm) pumps a PPLN
(Periodically Poled Lithium Niobate) nonlinear waveguide that creates pairs of photons through the process of spontaneous parametric
down-conversion. After the waveguide, a Silicon filter (F) blocks all the remaining light at 785.2 nm and the created photon pairs are
coupled into an optical fiber. Two circulators and two fiber Bragg gratings (FBG) separate the pairs according to their wavelength.
The first FBG reflects only the photons at 1573.0 nm ($\triangle\lambda$=1.0 nm) and allows for the rest of the photons to be transmitted, while
the second FBG reflects the photons at 1567.8 nm ($\triangle\lambda$=1.0 nm). The rest of the photons are not transmitted. The photons are sent
to their respective receiving stations through standard communication optical fibers. Although the photons wavelength was chosen to be very
close to the third telecommunication window at 1550 nm, there was still around 8 dB of losses in each of the fibers linking the source with
the detectors, concentrated mainly at the connectors between different fibers.
\begin{center}
\includegraphics[width=1.0\linewidth]{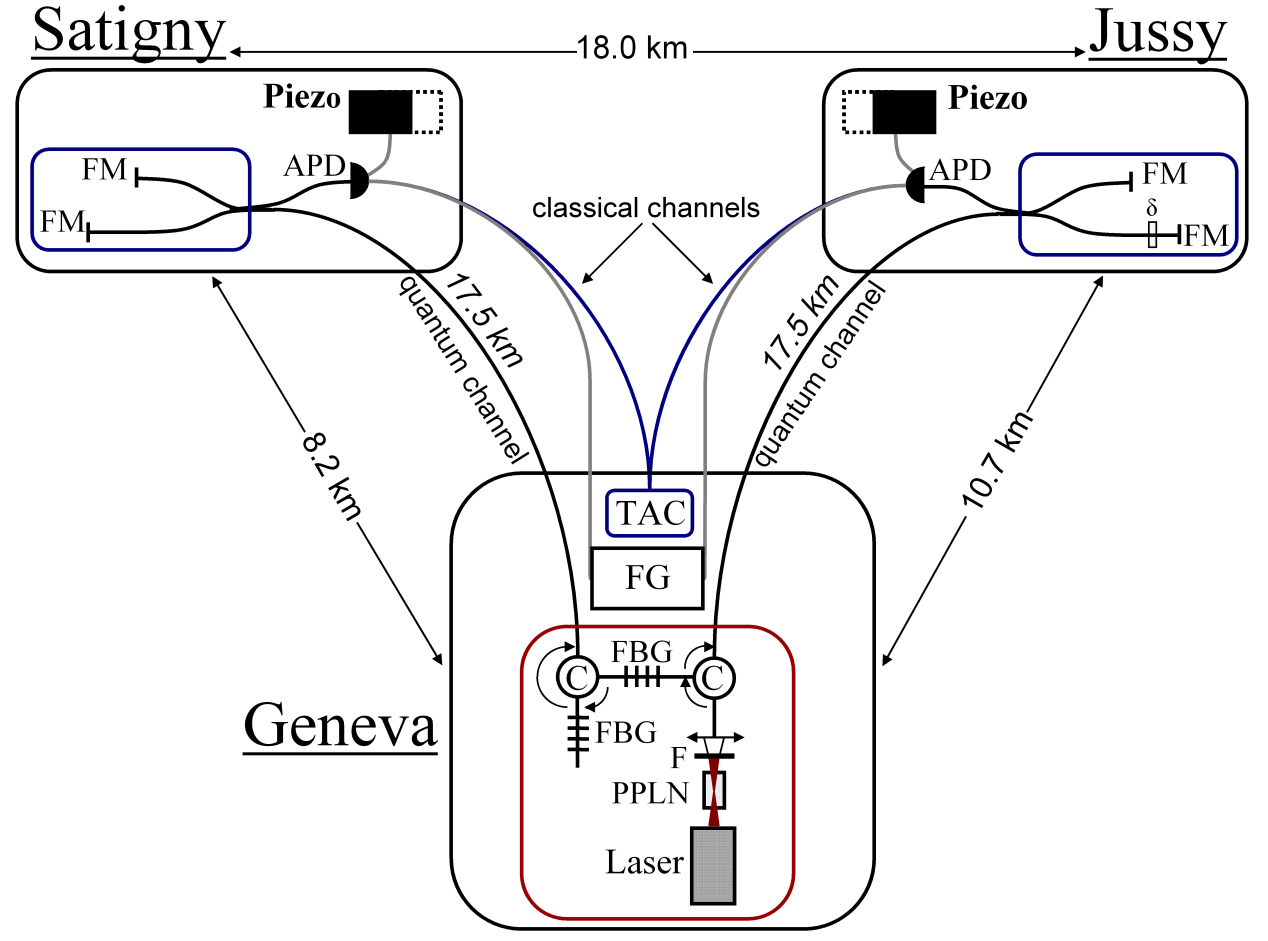}
\end{center}
{\small Fig. 4. Experimental setup. See text for a detailed description.}\\

The geographical layout of the experiment is such that the source is situated in Geneva and sends the pairs to two receiving stations situated in two villages (Satigny
and Jussy) in the Geneva region, at 8.2 and 10.7 km, respectively. The direct distance between them is 18.0 km. At each receiving station, the
photons pass through a Michelson interferometer with a long arm and a short arm. The path-length difference is 1.3 ns and is the same in both
interferometers. It is also smaller than the coherence length of the pump laser, so an entangled state can be detected when both photons
pass through the short (and long) arms. Because the photons are entangled, the probability that pairs of photons choose the same output port can be
affected by changing the phase in either interferometer. This will produce interference fringes in the coincidence count when the phase is
scanned. To scan the phase, the temperature of one of the interferometers is changed while the other is left stable. To compensate for
birefringence effects in the arms of the interferometers (i.e. to stabilize polarization), Faraday mirrors (FM) are used \cite{Mar}.

After passing through the interferometers, the photons are detected by single-photon InGaAs avalanche photodiodes (APDs)
(id Quantique, id200). The photodiodes are operated in the gated mode with a repetition frequency of 1 MHz and a gate width
of 100 ns. The quantum efficiency is $10\%$ and the dead time is $10\ \mu s$. They are triggered in a synchronized way using the same signal
sent out from Geneva through other optical fibers. This greatly improves the number of coincidences per unit of time.

Each time the single-photon APDs detect an event, a classical optical signal is sent back to Geneva, where it is detected by a p-i-n photodiode.
For this, we used the same fibers that were used to send the trigger signal to the APDs. The lasers at both ends of the fibers had different
wavelengths (1550 nm and 1310 nm) and Wavelength Division Multiplexers (WDM) were used to separate the signals. The detected events are sent
to a time-to-amplitude-converter (TAC) that takes one of the signals as start and the other as the stop, measuring the difference in their
arrival times. Coincidences in the arrival times between events coming from different detectors indicate that those photons passed either
through the short-short or the long-long paths in the interferometers. Using a discriminator with a narrow temporal window, the other two non-interfering
possibilities (photons that passed through different arms $-$ short-long or long-short paths) can be discarded.

We monitored the single photon count rates for each detector and the coincidence rate while scanning the phase $\delta$ in one of the
interferometers. We decrease the temperature of one interferometer slowly and regularly from $40^\circ C$ to $21^\circ C$ during a
period of several hours. The coherence length of the single photons was $2.5$ mm, 2 orders of magnitude smaller than the path-length
difference in the interferometers (267 mm), so there was no single photon interference, and no phase-dependent variations in the single rates
were observed. On the contrary, the coincidence rate showed a sinusoidal oscillation dependent of the phase change in the interferometers.
The single count rates were continuously controlled and found constant around 5.0 and 4.1 kHz, including 0.7 and 1.1 kHz of dark counts, for the detectors at Satigny and Jussy,
respectively. A discriminator window of 600 ps placed around the coincidence peak gave us an average coincidence rate of 33 coinc./min.

The Bell inequalities set an upper bound for correlations between particles that can be described by local theories.
One of the most frequently used forms is the Clauser-Horne-Shimony-Holt (CHSH) Bell inequality \cite{CHSH}, which has a Bell parameter
\begin{equation}
S = |E(d_1, d_2) + E(d_1, d_2') + E(d_1', d_2) - E(d_1', d_2')| \leq 2 \nonumber
\end{equation}
where $E(d_1, d_2)$ are the correlation coefficients and $d_1$, $d_2$ are values for the phase in the interferometers. Quantum mechanics predicts a
maximum value of $S=2\sqrt{2}$. If the correlation coefficient E is described by a sinusoidal function like $E=Vcos(\delta)$ where $\delta$
is the relative phase in the interferometers and V is the visibility, the parameter S becomes $S=2\sqrt{2}V$. This implies that if the
visibility is $V\geq 1/\sqrt{2}$ the correlations between detected photons is nonlocal.

We are interested in the best visibility value obtained over a period of a few fringes. To obtain an optimal visibility,
it is important to have less than 0.1 pairs/time-window in order to reduce the probability of having a double pair. The number of photon pairs
within the FBGs 1-nm bandwidth was $0.07$ pairs per 600 ps time-window. The raw data yields a visibility of $V_{raw}=(90.5\pm1.5)\%$ (see Fig. 5)
leading to $S_{raw} = 2.56\pm 0.04$, surpassing the limit given by the Bell inequalities by 13 standard deviations $(\sigma)$.
We can conclude that the correlations between the photons remain well above the local limit even when the gravitational field is being
modified by the displacement of the masses.

\begin{center}
\includegraphics[width=0.9\linewidth]{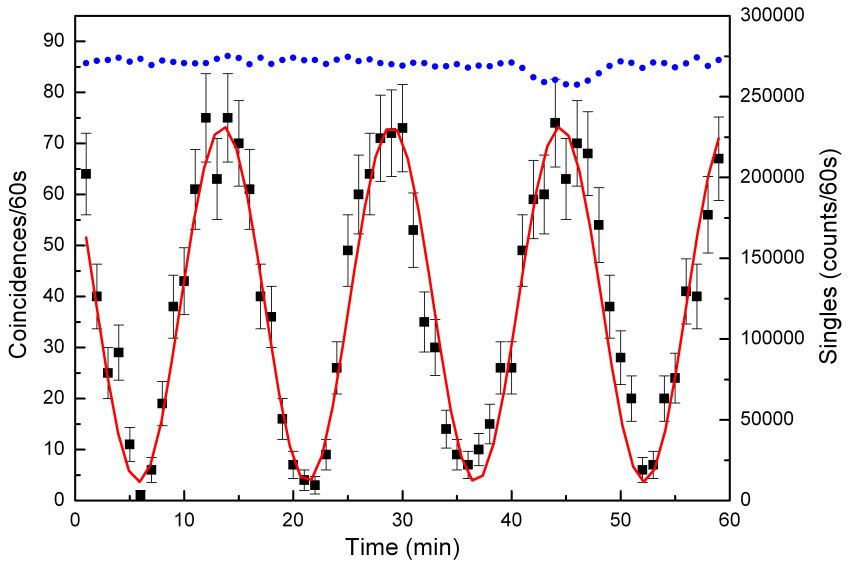}
\end{center}
{\small Fig. 5. Singles (dots) and coincidence counts (squares) per 60 s as a function of time while the phase $\delta$ in the Jussy interferometer
(see Fig. 4) was being scanned. A best fit with a sinusoidal function yields a visibility of $V_{raw}=(90.5 \pm 1.5)\%$. Error bars
represent the square root of the number of coincidences at each point.}\\

Sometimes an avalanche in one of the APDs is set off but without any photon.
If such a false detection happens at almost the same time in both APDs or if it happens when one true photon arrives at the other
APD, this leads to an accidental coincidence. The number of accidental coincidences will not oscillate with the scannning of the phase
but will always remain around the same value, reducing the visibility. The number of accidental coincidences was 2.5 coinc./min.
If we substract the accidentals from the total number of coincidences, the visibility climbs to $V_{net}=(96.7\pm1.4)\%$ leading
to $S_{net} = 2.74\pm 0.04$, violating the Bell inequalities by 18$\sigma$.

In conclusion, we have performed an experimental test of the Bell inequality with space-like separation large enough to include a hypothetical
delay of the quantum state reduction until a macroscopic mass has significantly moved, as advocated by Penrose and Di\'osi. Indeed, in the
reported experiment each detection event triggers the application of a step voltage that expands a piezo actuator and displaces a mirror.
The time of collapse of the mirror plus the time it takes to move it is shorter than the time the light needs to travel the distance
between the receiving stations. In addition, this distance (18 km) sets a new record for Bell experiments with an independent source located in
the middle. Let us emphasize that under the assumption that a quantum measurement is finished only once a gravity-induced state reduction has occurred, none of the many former Bell experiments involve
space-like separation, that is space-like separation from the time the particle (here photons) enter their measuring apparatuses (here interferometers) until the time the
measurement is finished. In this sense, our experiment is the first one with true space-like separation. The results confirm the nonlocal
nature of quantum correlations.

We acknowledge technical support by J-D. Gautier and C. Barreiro. The access to the telecommunication
network was gracefully provided by Swisscom. This work was supported by the Swiss NCCR Quantum Photonics and the EU project QAP.
\begin{center}
\line(1,0){150}
\end{center}
\end{document}